\documentclass[showkeys,preprint,showpacs,preprintnumbers,amsmath,amssymb,aps,superscriptaddress,prl]{revtex4}

\usepackage{graphicx}
\usepackage{dcolumn}
\usepackage{lscape}
\usepackage{bm}
\usepackage{latexsym,epsfig}

\newcommand{\simge}{\hspace*{0.2em}\raisebox{0.5ex}{$>$}
     \hspace{-0.8em}\raisebox{-0.3em}{$\sim$}\hspace*{0.2em}}
\newcommand{\simle}{\hspace*{0.2em}\raisebox{0.5ex}{$<$}
     \hspace{-0.8em}\raisebox{-0.3em}{$\sim$}\hspace*{0.2em}}
\newcommand{\ep}{\epsilon}

\newcommand{\al}{\alpha}
\newcommand{\bt}{\beta}
\newcommand{\g}{\gamma}

\newcommand{\dt}{\delta}
\newcommand{\si}{\sigma}
\newcommand{\la}{\lambda}
\newcommand{\simu}{\sigma^{\mu\nu}}
\newcommand{\Fmu}{F_{\mu\nu}}
\newcommand{\Gmu}{G^a_{\mu\nu}}

\newcommand{\slashT}{\slash\hspace{-0.4em}T}
\newcommand{\slashP}{\slash\hspace{-0.6em}P}
\newcommand{\slashPsub}{\slash\hspace{-0.5em}P}
\newcommand{\slashPT}{\slash\hspace{-0.6em}P\slash\hspace{-0.5em}T}

\newcommand{\qb}{\bar q}
\newcommand{\Nb}{\bar N}
\newcommand{\Fp}{F_\pi}

\newcommand{\tb}{\bar \theta}

\newcommand{\mpi}{m_{\pi}}
\newcommand{\MQCD}{M_{\mathrm{QCD}}}

\newcommand{\Or}{\mathcal O}

\newcommand{\dslash}[1]{#1 \llap{/\kern-0.5pt}}
\newcommand{\Dslash}[1]{#1 \llap{/\kern+1.2pt}}
\newcommand{\DDslash}[1]{#1 \llap{/\kern+2.3pt}}
\newcommand{\dslashh}[1]{#1 \llap{/\kern+1pt}}
\newcommand{\abs}[1]{|#1|}

\newcommand{\boldtau}{\mbox{\boldmath $\tau$}}
\newcommand{\boldpi}{\mbox{\boldmath $\pi$}}

\begin{document}

\title{Parity- and Time-Reversal-Violating Form Factors of the Deuteron}

\author{J. de Vries}
\affiliation{KVI, Theory Group, University of Groningen,
 9747 AA Groningen, The Netherlands}

\author{E. Mereghetti}
\affiliation{Department of Physics, University of Arizona,
 Tucson, AZ 85721, USA}

\author{R. G. E. Timmermans}
\affiliation{KVI, Theory Group, University of Groningen,
 9747 AA Groningen, The Netherlands}

\author{U. van Kolck}
\affiliation{Department of Physics, University of Arizona,
 Tucson, AZ 85721, USA}

\date{\today}

\begin{abstract}
We calculate the electric-dipole and magnetic-quadrupole form factors of the 
deuteron that arise as a low-energy manifestation of parity and time-reversal
violation in quark-gluon interactions. 
We consider the 
QCD vacuum angle 
and the dimension-six operators that originate from physics 
beyond the Standard Model: 
the quark electric and chromo-electric dipole moments, 
and the gluon chromo-electric dipole moment. 
Within the framework of two-flavor chiral perturbation theory, 
we show that in combination with the nucleon electric dipole moment, 
the deuteron moments would allow an identification of the dominant 
source(s) of  
symmetry violation.
\end{abstract}
\pacs{13.40.Gp, 11.30.Er }

\maketitle
Permanent electric dipole moments (EDMs) of particles, nuclei, atoms, and
molecules violate both parity ($P$) and time-reversal ($T$), or equivalently 
$C\!P$, invariance \cite{Pos05}.
Since $C\!P$ violation due to quark mixing in the electroweak sector of the 
Standard Model (SM) seems insufficient to explain the observed 
matter-antimatter asymmetry in the universe \cite{Coh93}
and predicts immeasurably small values for the EDMs of nucleons and nuclei 
\cite{Khr82}, searches for nonzero EDMs are an excellent probe for new sources
of  $C\!P$ violation. 
Experiments with ultracold neutrons are in preparation that 
aim to improve the bound on the neutron EDM~\cite{Bak06} by two orders of 
magnitude~\cite{nEDM}. Moreover, plans exist to measure the EDMs of light ions,
in particular the proton and the deuteron, in storage ring experiments at 
similar levels of accuracy~\cite{Far04}. The current generation of ongoing 
and planned EDM experiments probes the same energy scales as the LHC, 
and there are strong expectations that nonzero results will soon be found.

An outstanding theoretical issue is to identify the fundamental  
$C\!P$-violating
mechanisms, and in particular to relate a positive signal in the EDM
experiments to the $P$- and $T$-violating ($\slashPT$) sources at the 
quark-gluon level. 
The SM contains the $\slashPT$ QCD $\tb$ term \cite{'tHooft:1976up}, 
which has dimension four and would  be expected to give the main contribution 
to hadronic $\slashPT$.
However, since the experimental upper limit on the neutron EDM  constrains
$\tb$ to be unnaturally small~\cite{Cre79}, 
$\tb \simle 10^{-10}$, 
possible contributions from higher-dimensional $\slashPT$ sources can be 
relevant, 
or even dominant.
These higher-dimensional operators have their origin beyond the SM, 
in an ultraviolet complete theory at a high energy scale $M_{\slashT}$,
for example a supersymmetric version of the SM \cite{RamseyMusolf:2006vr}.
The first such effective $\slashPT$ operators  one encounters have dimension
six~\cite{Ruj91,Wei89}, {\it viz.} the quark EDM (qEDM) and the quark and gluon
chromo-electric dipole moments (qCEDM and gCEDM).
We show that in combination with the nucleon EDM, a measurement of the deuteron
$\slashPT$ 
form factors (FFs) would allow the disentanglement of these $\slashPT$ sources.

The difficulty in calculating such low-energy observables stems 
from the breakdown of perturbation theory in the 
QCD coupling constant below the characteristic QCD scale 
$\MQCD\sim 2\pi F_\pi\simeq 1.2$ GeV, with $F_{\pi}=185$ MeV 
the pion decay constant.
For processes at momenta $Q\sim m_\pi$, the mass of the lightest hadron,
the pion, we can nevertheless express observables
in a controlled expansion in powers of $Q$ 
using chiral perturbation theory (ChPT) \cite{Wei79}
with the two lightest quark flavors $u$ and $d$.
This effective field theory (EFT) 
involving pions, nucleons and photons
correctly incorporates 
the (approximate) symmetries of QCD, in particular 
the spontaneously and explicitly broken $SO(4)$ chiral symmetry, 
and describes low-energy physics 
in a model-independent way. 
All effective hadronic interactions that transform under symmetries as 
terms in the QCD Lagrangian are allowed,
each one being associated with a parameter, or low-energy constant (LEC), 
which can be estimated using naive dimensional analysis (NDA)~\cite{NDA,Wei89}.
Since the $\slashPT$ sources break chiral symmetry in different
ways \cite{Mer10}, they can in principle be distinguished by the
form and expected strength of their hadronic interactions.

The nucleon EDFF partially reflects the $\slashPT$ source
at the quark-gluon level~\cite{Hoc05,Vri11a,Mer11}: 
a measurement of the nucleon EDM 
and its Schiff moment~\cite{Tho94} could distinguish between 
$\slashPT$
originating from the $\tb$ term or qCEDM on the one hand, and the qEDM or 
gCEDM on the other. 
Since also the deuteron can be analyzed with firm theoretical tools,
we focus here on its $\slashPT$ electromagnetic FFs 
which, part from small relativistic corrections, 
are defined 
from 
the $\slashPT$ part of the electromagnetic current,
$J^\mu_{\slashPsub\slashT}$, 
by
\begin{eqnarray}
\langle\vec p^{\, \prime},j|J^0_{\slashPsub\slashT}|\vec p,i\rangle &=& -
\ep^{ijl}q^l 
F_D(\vec q^{\, 2}),
\\
\langle\vec p^{\, \prime},j|J^k_{\slashPsub\slashT}|\vec p,i\rangle &=&-
\ep^{mnl} q^l
\left[
\dt^{mi}\dt^{nj}K^k 
\frac{F_D(\vec q^{\, 2})}{m_d}
-\frac{1}{4}\dt^{mk}(\dt^{ni}q^j+\dt^{nj}q^i)
F_M(\vec q^{\, 2})
\right],
\label{formfactors}
\end{eqnarray}
where 
$|\vec p,i\rangle$ denotes a deuteron state of momentum $\vec p$ and
polarization 
$\delta^\mu_i$ in the rest frame, normalized so that 
$\langle\vec p^{\, \prime},j|\vec p,i\rangle = 
\sqrt{1+\vec p\,^2/m_d^2}(2\pi)^3 \delta^{(3)}(\vec q\,) 
\delta_{ij}$, 
$\vec q=\vec p-\vec p^{\, \prime}$ is the photon momentum,
$\vec K=(\vec p^{\, \prime}+\vec p)/2$,
and $m_d=2m_N -\gamma^2/m_N +\ldots$ is the deuteron mass in terms
of the nucleon mass $m_N$ and the binding momentum $\gamma$.
We show that a combination of the deuteron EDM, $d_d=F_D(0)$, and
nucleon EDM can separate the qCEDM from the other $\slashPT$ sources,
and that a measurement of the magnetic quadrupole 
moment (MQM), $\mathcal M_d = F_M(0)$, is
sensitive to the $\tb$ term.

$P$- and $T$-conserving ($PT$) pion interactions
are relatively weak, because they proceed through derivatives
when originating
from the couplings of quarks and gluons,
which are chiral-symmetric, or through powers of the small 
chiral-symmetry breaking parameters, the quark masses 
$m_{u,d}\sim \bar{m} ={\cal O}(m_\pi^2/\MQCD)$ and 
the proton charge $-e=\sqrt{4\pi\alpha_{em}}$.
In the one-nucleon sector 
the expansion is in powers of $Q/\MQCD$, but
subtleties in the power counting arise in the two-nucleon ($N\!N$) 
sector \cite{review2}, 
where one has to accommodate the large scattering lengths in the 
$^1S_0$ and $^3S_1$ waves, and the related presence of the unnaturally
shallow bound state in the $^3S_1$ wave, the deuteron. 
Such fine-tuning can be incorporated if the 
$N\!N$ LECs are assigned a scaling~\cite{KSW}
with inverse powers of the small scale
$\gamma\simeq 45$ MeV.
Moreover, the strength of pion exchange among nucleons 
is set by $M_{N\!N} \sim 4\pi \Fp^2/m_N \simeq 450$ MeV,
and for momenta
$Q\simle M_{N\!N}$
pions can be treated perturbatively~\cite{KSW},
the deuteron arising when
the leading $N\!N$ contact interaction is summed to all orders. 
$N\!N$ observables are amenable to
an additional expansion in powers of $Q/M_{N\!N}$, 
although the breakdown scale of this expansion in scattering \cite{KSW}
suggests that $M_{N\!N}$ is smaller than guessed above by a factor of 2 or 3.
This scheme was used to
describe the low-energy properties of the deuteron, 
in particular its 
$PT$ (charge and magnetic dipole) and $\slashP T$
(anapole) FFs \cite{Kap99}.
We calculate here the  
$\slashPT$ FFs 
in leading order (LO) for the first time.

At the level of the quark field $q=(u \; d)^T$ and
the photon and gluon field strengths $\Fmu$ and $G_{\mu \nu}^a$,
the lowest-dimension $\slashPT$ sources are
\begin{eqnarray}
{\cal L}_{\slashPsub\slashT} &=& m_\star \bar{\theta} \; \bar{q}i\gamma_{5}q
   -\frac{i}{2}\qb \left(d_0+d_3 \tau_3\right)\simu \g_5 q \; \Fmu 
\nonumber\\
 &&-\frac{i}{2}\qb \left(\tilde{d}_0+\tilde{d}_3 \tau_3\right)
                   \simu\g_5\la^a q \; \Gmu
+\frac{d_W}{6}\ep^{\mu\nu\la\si}f^{abc} 
     G^a_{\mu \rho}G_\nu^{b,\rho}G^c_{\la \si},
\label{eq:dim6}
\end{eqnarray}
in terms of the Pauli isospin matrices $\boldtau$,
the Dirac spin matrices $\gamma_5$ and $\sigma^{\mu\nu}$,
and the Gell-Mann color matrices $\la^a$ and structure constants $f^{abc}$.
The first term, with $m_\star=m_um_d/(m_u+m_d)\simeq \bar m/2$,
incorporates the QCD angle $\bar \theta$ \cite{'tHooft:1976up, Cre79}.
The second (third) term represents the isoscalar $d_0$ ($\tilde d_0$) 
and isovector $d_3$ ($\tilde d_3$) components of the qEDM (qCEDM). 
In the last term, $d_W$ is the gCEDM \cite{Wei89}.
Because of electroweak $SU(2)\times U(1)$ gauge symmetry,
the qEDM and qCEDM 
are proportional to the vacuum expectation value of the Higgs field, 
and therefore have effective dimension six \cite{Ruj91},
being suppressed by two powers of $M_{\slashT}$.
We write \cite{Vri11a}
\begin{eqnarray}
d_i =\Or\!\left(\frac{e\delta \bar m}{M^2_{\slashT}}\right),\,\,\,\,\,\,
\tilde d_i =\Or\!\left(\frac{4\pi \tilde\delta\bar m}{M^2_{\slashT}}\right),\,\,\,\,\,\,
d_W =\Or\!\left(\frac{4\pi w}{M^2_{\slashT}}\right),
\end{eqnarray}
where $\delta$, $\tilde \delta$, and $w$ are dimensionless parameters 
that depend on the mechanisms of electroweak and $PT$ breaking, 
and on the running 
from the electroweak scale $M_W$
to low energies;  their sizes can be calculated in specific
high-energy models 
in terms of coupling constants and 
complex phases 
\cite{Pos05, RamseyMusolf:2006vr}.
Our approach is limited to low energies, where the contributions
associated with heavier quarks can be buried in the LECs.
Effects of higher-dimension
$\slashPT$ sources should
be suppressed by $M_W^2/M_{\slashT}^2$.

While all interactions in Eq. (\ref{eq:dim6}) break $P$ and $T$, 
each transforms under $SO(4)$ in a characteristic way 
\cite{Hoc05,Mer10,Vri11a,Mer11}.
The $\bar\theta$ term is the fourth component of the same $SO(4)$
vector $P=(\bar{q}\boldtau q,\bar{q} i\gamma_5 q)$
that leads to isospin breaking~\cite{Hoc05,Mer10,Mer11},
and thus generates EFT 
interactions that transform as $\slashPT$ fourth components 
of $SO(4)$ vectors made out of hadronic fields, with
coefficients related to those of $PT$ interactions. 
Similarly, the qCEDM and qEDM both break chiral symmetry as combinations 
of fourth and third components of two other $SO(4)$ vectors 
\cite{Vri11a,Mer11}. The gCEDM does not break chiral symmetry.
As in the $PT$ case, we use NDA 
to estimate the strength of the effective interactions. 

The relevant $\slashPT$ Lagrangian \cite{Mer10}
in terms of the pion field $\boldpi$ and  the heavy-nucleon field
$N=(p \; n)^T$ of velocity $v^\mu$ and spin $S^\mu$ is
\begin{eqnarray}
{\cal L}_{\slashPsub\slashT}&=&
-\frac{1}{F_\pi} 
 \bar{N}
 \left(\bar{g}_{0}\boldpi\cdot\boldtau+\bar{g}_{1} \pi_3 \right) N
 +2\bar{d}_0 \,\Nb S^\mu  N\, v^\nu \Fmu \nonumber\\
&&
 +\frac{\bar c_\pi}{\Fp} \ep^{\mu \nu \al \bt}  v_\al 
 \, \Nb S_\bt \boldpi\cdot\boldtau N  \,  \Fmu+\bar C_0 \left[ \Nb\!N \, \partial^\mu\left(\Nb S_\mu N\right)
- \Nb \boldtau N\cdot \partial^\mu\left(\Nb \boldtau S_\mu N\right)\right]\nonumber\\
&&
 +\bar M \ep^{\mu \nu \al \bt}  v_\al 
 \, \Nb S_\bt N \, \Nb S_\lambda N \, \partial^\lambda \Fmu+\ldots,
\label{LslashT}
\end{eqnarray}
where $\ep^{0123}=1$,  
\begin{equation}
\bar{g}_{0}= 
\Or\!\left(\bar{\theta}\frac{m_\pi^2}{\MQCD},
              \tilde{\delta}\frac{m_\pi^2\MQCD}{M_{\slashT}^2}\right),\,\,\,\,\,\,
\bar{g}_{1}= 
\Or\!\left(\tilde{\delta}\frac{m_\pi^2\MQCD}{M_{\slashT}^2}\right)
\label{NDA1}
\end{equation}
are $\slashPT$ pion-nucleon couplings,
\begin{eqnarray}
\bar{d}_{0}= 
\Or\!\left(e\bar{\theta}\frac{m_\pi^2}{\MQCD^3},
             e\delta\frac{m_\pi^2}{\MQCD M_{\slashT}^2},
             e w \frac{\MQCD}{M_{\slashT}^2}\right)
\label{NDA2}
\end{eqnarray}
contributes to the short-range isoscalar nucleon EDM,
\begin{equation}
\bar c_{\pi} = 
\Or\!\left(e\delta\frac{m_\pi^2}{\MQCD M_{\slashT}^2}\right)
\label{NDA5}
\end{equation}
is a $\slashPT$ pion-nucleon-photon interaction,
\begin{equation}
\bar C_{0} = 
\Or\!\left(w\frac{4\pi}{m_N \gamma}\frac{\MQCD}{M_{\slashT}^2}\right)
\label{NDA3}
\end{equation}
is the leading $\slashPT$ $N\! N$ contact LEC,
\begin{eqnarray}
&&\bar M= 
\Or\!\left(e\delta \frac{4\pi}{m_N \gamma^2}
                      \frac{m_\pi^2 }{M_{N\!N}\MQCD M_{\slashT}^2}\right)
\label{NDA4}
\end{eqnarray}
parametrizes short-distance $N\! N$ $\slashPT$ currents,
and ``$\ldots$'' stand for terms that only contribute
to the FFs at higher orders. 
For $\tb$, the link with isospin violation \cite{Mer10}
implies $\bar g_0 \simeq \delta m_N (m_d+m_u) \tb/2(m_d-m_u)
\simeq 2.8 \, \tb$ MeV, using lattice QCD input \cite{latticedeltamN}
for the quark-mass piece of the nucleon mass difference $\delta m_N$.

\begin{figure}[t]
\centering
\includegraphics[scale = 0.6]{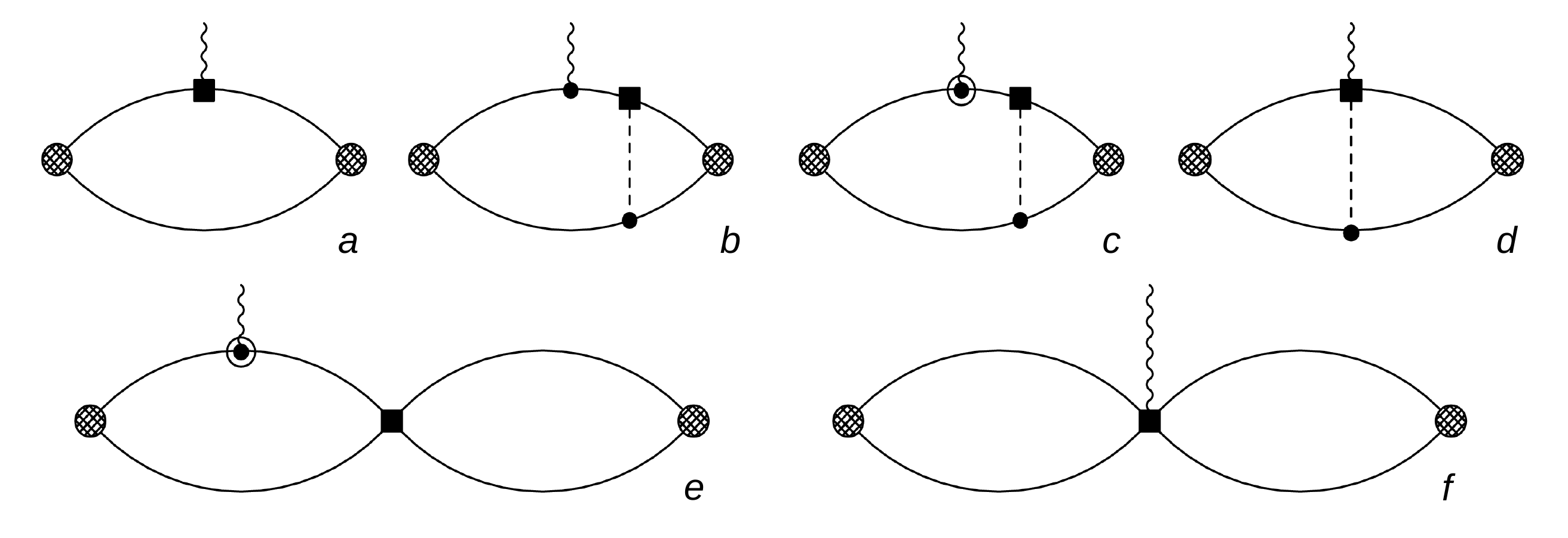}
\caption{LO diagrams for the deuteron 
EDFF ($a$, $b$) and MQFF ($c$, $d$, $e$, $f$).
Solid, dashed, and wavy lines represent nucleons, pions, and photons.
A square marks $\slashPT$, and the other vertices $PT$ interactions:
leading (filled circles) and subleading (circled circles). 
The hatched vertex represents the deuteron state.
Only one topology per diagram is shown.}
\label{graphs}
\end{figure}

The calculation of the EDFF and MQFF involves at LO the diagrams  of
Fig. \ref{graphs}, where the squares denote interactions 
from ${\cal L}_{\slashPsub\slashT}$.
The circles denote well-known $PT$ interactions, 
see {\it e.g.} Ref. \cite{review1}.
The pion-nucleon vertex is the standard axial-vector coupling, $g_{A}=1.27$.
The photon vertex denoted by a filled circle is the
coupling to the charge $e$, and that denoted by a circled circle
is the magnetic coupling parametrized by the anomalous magnetic moments,
the isoscalar $\kappa_0=-0.12$ and the isovector $\kappa_1=3.71$. 
The hatched circles denote deuteron states \cite{Kap99} obtained from the
iteration of the leading $N\! N$ contact interaction,
whose LEC can be eliminated in favor of $\gamma$.
We use dimensional regularization with power-divergence subtraction
\cite{KSW} at a renormalization scale $\mu$.
Our results depend on the ratio $\xi=\g/\mpi$ and on 
three functions of the momentum in the ratio $x=\abs{\vec q\,}/4\g $:
\begin{equation}
F_1(x)=\arctan (x)/x,
\end{equation}
which originates in a bubble with one photon coupling and appears also
in the charge FF \cite{Kap99},
and two complicated functions that result from two-loop diagrams
with a pion propagator, which can be expanded as
\begin{equation}
F_2(x) = 1
-x^2 \frac{10+65\xi+144\xi^2+72\xi^3}{30 (1+\xi) (1+2\xi)^2} 
+ {\cal O}(x^4),
\end{equation}
\begin{equation}
F_3(x) = 1
-x^2 \frac{\xi^2 (12+8\xi)}{5 (1-2\xi) (1+2\xi)^2} 
+ {\cal O}(x^4) .
\end{equation}
The scale of momentum variation 
is set by $4\gamma$.

The LO deuteron EDFF is due to diagrams $a$ and $b$,
\begin{equation}
F_{D}
(\vec q^{\, 2})
= 
2\bar{d}_{0} F_1(x)
-\frac{e g_A \bar g_1 m_N}{6 \pi \Fp^2 \mpi} 
\frac{1+\xi}{(1+2\xi)^2} F_2(x),
\label{onebodyEDFF}
\end{equation}
where the first term is dominant for $\tb$, qEDM, and gCEDM,
and the second one for qCEDM.
The LO MQFF 
comes from diagrams $c$, $d$, $e$, and $f$,
\begin{eqnarray}
F_{M}
(\vec q^{\, 2})
&=&\frac{e (1+\kappa_0)}{2\pi}
  (\mu -\g)\bar C_{0}
F_1(x)\nonumber\\
&&+ \frac{e g_A }{2\pi\Fp^2 \mpi}
\left[\bar g_0 (1+\kappa_0)+\frac{\bar g_1}{3}(1+\kappa_1)\right]
 \frac{1+\xi}{(1+2\xi)^2} F_2(x)
  \nonumber\\
&&+\frac{2 \g }{\pi}(\mu -\g)^2 \bar M +\frac{g_A \bar c_\pi \g}{\pi\Fp^2}\left(\frac{1-2\xi}{1+2\xi}F_3(x)
+2\ln\frac{\mu/m_\pi}{1+2\xi}\right),   
\end{eqnarray} 
where at this order $\bar g_0$ originates from $\tb$ and qCEDM, 
$\bar g_1$ from qCEDM only, $\bar C_0$ from gCEDM,
and $\bar M$ and $\bar c_\pi$ from qEDM. 

We can now discuss the implications of the various $\slashPT$ sources for
the deuteron EDFF and MQFF.  In Table~\ref{table1} we list the orders of
magnitude for the deuteron EDM, $d_d$, the ratio of deuteron-to-neutron EDMs,
$d_d/d_n$, and the ratio of the deuteron MQM and EDM, ${\cal M}_d/d_d$,
for the different $\slashPT$ sources.
Just as for $d_n$ \cite{Cre79,Hoc05,Vri11a}, a $d_d$ signal by itself could
be attributed to any source with a parameter of appropriate size. 
For $\tb$, qEDM, and gCEDM the deuteron EDFF is determined  
by the LO isoscalar nucleon EDM, and thus
well approximated by the sum of neutron and proton EDM.
For $\tb$ in particular, using the most important long-range contributions, 
which appear at NLO, as a lower bound for $\bar d_0$ \cite{Mer11,Ott10}, 
one finds $|d_d|\simge 2.8 \cdot 10^{-4}\,\tb$ $e\,$fm.
If, however, the dominant
$\!\slashPT$ source is the qCEDM, $d_d$
comes mainly from neutral-pion exchange. 
A measurement of $\abs{d_d}$ significantly larger than $\abs{d_n}$ 
would be indicative of a qCEDM. 
A null-measurement at the $10^{-16}$  $e\,$fm level \cite{Far04} would 
strengthen  
the bounds from the neutron
\cite{Vri11a} 
to $\tb\simle 3\cdot10^{-13}$ and 
${\tilde \delta}, w, \, 3\cdot10^{-2} \delta 
\simle (M_{\slashT}/3\cdot  10^{7} \; {\rm GeV})^{2}$.
More quantitative statements could be made with lattice-QCD
calculations of the EFT LECs.

\begin{table}[t]
\caption{Orders of magnitude for the deuteron EDM 
(in units of 
$em_d^{-1}$), the ratio of deuteron-to-neutron EDMs, 
and the ratio of the deuteron MQM and EDM (in units of 
$m_d^{-1}$), 
for $\slashPT$ sources of effective dimension up to six.}
\begin{tabular}{c|cccc} 
\hline\hline
Source & $\tb$ & qCEDM & qEDM & gCEDM \\ 
\hline
$m_d \, d_d/e$  & 
$\tb \frac{\mpi^2}{\MQCD^2}$ & 
${\tilde \delta}\frac{\mpi\MQCD^2}{M_{N\!N}M_{\slashT}^2}$ &
$\delta \frac{\mpi^2}{M_{\slashT}^2}$ & 
$w \frac{\MQCD^2}{M_{\slashT}^2}$\\
$d_d/d_n$ & 
$1$ & 
$\frac{\MQCD^2}{\mpi M_{N\!N}}$ & 
$1$ & 
$1$ \\
$m_d \,\mathcal M_d /d_d$ & 
$\frac{\MQCD^2}{\mpi M_{N\!N}}$ & 
$1$ &
$\frac{\g}{M_{N\!N}}$ &
 $1$ \\ 
\hline\hline
\end{tabular}
\label{table1}
\end{table}

Additional information comes from the ratio ${\cal M}_d/d_d$. 
For $\tb$, $m_d\abs{{\cal M}_d}$ is expected to be larger than $\abs{d_d}$,  
whereas for
the dimension-six sources we expect $m_d\abs{{\cal M}_d}$ to be of similar size
or 
somewhat smaller than $\abs{d_d}$. 
For $\tb$, ${\cal M}_d$ is determined by pion exchange, and
we can again use the link with isospin violation \cite{Mer10} 
to find ${\cal M}_d 
\simeq 2.0 \cdot 10^{-3}\,\tb$ $e\,$ fm$^2$. 
An upper bound on ${\cal M}_d$ can therefore constrain $\tb$ 
without  relying on an estimate of short-range physics via the size of the
chiral log, which is necessary when using $d_n$ \cite{Cre79}.
Moreover, if $m_d\abs{{\cal M}_d}$ is found to be much smaller than 
$\abs{d_d}$, the source would likely be qEDM. 
This shows that  
a measurement of ${\cal M}_d$, 
in addition to $d_n$ and $d_d$, would be very valuable,
and as a consequence its feasibility is beginning to be investigated
\cite{underwater}.

The deuteron EDM and MQM were calculated previously in 
Ref.~\cite{Khr99}.
Since these calculations did not use the chiral properties of the fundamental 
$\slashPT$ sources, the $\slashPT$ pion-nucleon interactions were assumed 
to be all of the same size. When the dominant 
source is the qCEDM, their results agree with ours.
The advantage of our EFT framework is that it has a 
direct link to QCD by exploiting the chiral
properties of the $\slashPT$ dimension-four and -six operators. 
This is demonstrated by the $\bar g_2 \Nb \pi_3 \tau_3 N$ interaction 
used in many previous calculations, which  due to its chiral properties
only comes in at higher order for all $\slashPT$ sources \cite{Mer10}. 
Consequently, for the qCEDM, the ratio of $d_d$ to ${\cal M}_d$
depends at LO only on the ratio $\bar g_1/\bar g_0$, 
which can be measured independently: 
$\bar g_1$ could be inferred from $d_d$,
and $\bar g_0$ in principle from another observable,
such as the proton Schiff moment \cite{Vri11a} 
or the $^3$He EDM \cite{Ste08}. 
In addition, the power-counting
scheme allows a perturbative framework with analytical results that 
can be improved systematically.
Under the assumption that higher-order results are not afflicted by 
anomalously-large dimensionless factors, the relative error of
our results should be $\gamma/M_{N\!N}\sim 30$\%, as was explicitly verified 
for the charge FF \cite{Kap99}.
Our estimates for $d_d$ are consistent with those
from QCD sum rules \cite{Leb04}. 

In summary, we have investigated the leading-order, low-energy 
electric-dipole and magnetic-quadrupole 
form factors of the deuteron that result from the $\bar\theta$ angle,
the quark electric and chromo-electric dipole moments, and 
the gluon chromo-electric dipole moment. 
While for qCEDM we expect $\abs{d_d}$ to be larger 
than $\abs{d_n}$ by a factor $\Or(\MQCD^2/\mpi M_{N\!N})$,
for the other $\slashPT$ sources we have shown that $d_d$
is given by the sum of $d_n$ and $d_p$. 
Furthermore, the SM predicts $m_d\abs{{\cal M}_d}$ to be larger 
than $\abs{d_d}$, whereas beyond-the-SM physics 
prefers $m_d\abs{{\cal M}_d}$ smaller than, or of similar size as $\abs{d_d}$.
EDM and MQM measurements are therefore complementary.

\acknowledgments
We thank K. Jungmann, G. Onderwater, Y. Semertzidis, and E. Stephenson
for discussions and encouragement. This research was supported by the
Dutch Stichting FOM under programs 104 and 114 (JdV, RGET) and by the US
DOE under grants DE-FG02-06ER41449 (EM) and DE-FG02-04ER41338 (EM, UvK).

\end{document}